\documentclass[3p]{elsarticle}

\usepackage{amssymb}
\usepackage{amsthm}

\journal{JOI}

\begin{document}

\begin{frontmatter}



\title{Predicting the popularity of scientific publications by an age-based diffusion model}


\author{Yanbo Zhou}
\author{Qu Li}
\cortext[mycorrespondingauthor]{Corresponding author}
\ead[url]{liqu@zjut.edu.cn}
\author{Xuhua Yang}
\author{Hongbing cheng}

\address{College of Computer Science and Technology, Zhejiang University of Technology, Hangzhou, P. R. China}

\begin{abstract}
Predicting the popularity of scientific publications has attracted many attentions from various disciplines. In this paper, we focus on the popularity prediction problem of scientific papers, and propose an age-based diffusion (AD) model to identify which paper will receive more citations in the near future and will be popular. The AD model is a mimic of the attention diffusion process along the citation networks. The experimental study shows that the AD model can achieve better prediction accuracy than other network-based methods. For some newly published papers that have not accumulated many citations but will be popular in the near future, the AD model can substantially improve their rankings. This is really critical, because identifying the future highly cited papers from massive numbers of new papers published each month would provide very valuable references for researchers.

\end{abstract}




\end{frontmatter}


\section{Introduction}
With the increase of researches and the pressures that drive scholars to continuously produce publishable results\cite{plos2010}, the volume of scientific literatures is growing exponentially\cite{nature2016}. That brings out the information overload problem, and it's hard to retrieve and evaluate scientific literatures effectively. Measuring the scientific influence of scientific publications has been one of the focuses of scietometrics\cite{review2017pr}. Citation count and its variations are widely used to evaluate the influence of scientists and scientific publications\cite{review2015,review2016ji}, such as total citation count for papers and scientists, H-index\cite{hindex} for scientists, and impact factor\cite{if} for journals. The purpose of most of these metrics is to evaluate the historical influence of publications or scientists. Predicting the popularity of scientific publications and prestige of scientists is still a challenging issue which has attracted much attention from various disciplines. It can help researchers to find potential popular research directions, and can provide very valuable references for institutions to decide who they should hire or which project they should fund.

As the citation relationship of papers can be expressed as a directed network, many network-based models are used to evaluate the influence of scientific papers. According to preferential attachment theory, the future influence of papers is in proportion to the historical influence\cite{paperpa}. The linear extrapolation method based on preferential attachment mechanism has a strong bias to the old papers which have accumulated many citations, and it's hard to predict the future influence of newly published papers which have not yet get many citations. By using z-score to factor out the bias that caused by aging in preferential attachment\cite{newman2009}, some newly published papers that have not yet received many citations but will receive much attentions in the future can be detected\cite{newman2014}. Similar approach can be introduced to other methods that have an obvious bias by age\cite{medo2016ji} or fields\cite{medo2017ji}. In a network, the intrinsic fitness theory assumed that the probability of a node to attract new links is proportional to its intrinsic fitness value\cite{fitness1,fitness2}. Citation networks are temporal networks, and the fitness values of papers decay exponentially with time\cite{medoprl}. Considering the temporal decay effect, using a weighted citation count to predict the future popularity of papers can improve the prediction accuracy\cite{zhou2020}, and this method performs well in predicting the future popularity of newly published papers. A thorough study of the life-cycle of papers suggests that the citation rate of a paper increases in a few years after its publication, reaches a peak and then decreases rapidly\cite{adecay2015}. Considering the preferential attachment, fitness, and aging effects, the future citations of a paper can be predicted through establishing optimal fitting parameters\cite{science2013}.

Citation count based methods only use local information of citation networks to predict the popularity of papers. However, in real citation networks, nodes are heterogenous\cite{zhou2011}. The structures of citation networks should be analyzed to evaluate the influence of papers. PageRank is a process of random walk on a directed network, and it was initially introduced to measure the importance of web pages (\cite{pagerank,anatomy}). Citation network has common features with WWW, since they are both directed network and every incoming link corresponds to a positive rating. So Pagerank can naturally be used to citation networks to rank scientific papers\cite{sciPR1,sciPR2}. However, Pagerank dose not consider the aging effects in bibliometrics\cite{medoprl,adecay2015}. CiteRank\cite{citerank} (CR) is a more practical model for the diffusion along the citation network. It is an adaptation of the PageRank algorithm to citation networks, and it gives more weight to recent publications while starting a random work. The node scores of CiteRank can be used to predict the future popularity of papers. There are some other variants of PageRank that infuse time-awareness to evaluate the influence of papers\cite{TKDE2019}. In the time-aware methods, the importance of a citation depends not only on the importance of the citing paper, but also on the temporal features of citations.

In this paper, we focus on the popularity prediction problem of scientific papers, and propose an age-based diffusion (AD) model to predict the future popularity of scientific papers. The AD model considers the aging effects in bibliometrics, and employs a process analogous to random walk across citation networks. While distributing a paper's score to its references, an exponential decay mechanism is introduced and the probability of distributing a paper's score to its references decays with the iteration of steps. The experimental results show that the AD model can identify the future popular papers in their early life and improve the overall citation prediction accuracy. The rest of this paper is organized as follows: In Section 2 we introduce the data set we used, briefly review some benchmark methods and introduce the age-based diffusion (AD) model. In Section 3 we systematically evaluate the effectiveness of the proposed model in predicting citations of papers. Finally, we conclude the paper in section 4.

\section{Material and methods}
\subsection{Data Description}

We use the bibliographic data set provided by the American Physical Society (APS)\footnote{http://journals.aps.org/datasets} for analysis in this paper. The data set contains article metadata and citing article pairs of the corpus of Physical Review Letters, Physical Review, and Reviews of Modern Physics from 1893 through 2017. We pre-processed the date and removed the papers with incomplete information. And finally, there are 616316 papers with 7336550 citations among them. In the original data set, the publication date of each paper is recorded in the resolution of day. In this paper, we use time resolution of month to record papers' publication date. And for each paper, we used its publication month as its publication date. The citation relationship of papers can be expressed by a directed network. If paper $i$ cites paper $j$, there will be a link from paper $i$ to paper $j$ in the directed network.

\subsection{Related Metrics}

In this subsection, we will introduce three related network based methods that can be used to predict the popularity of scientific papers.

\subsubsection{PageRank}

PageRank (PR) is a process of random walk on a directed network, and can naturally be used to citation network to rank scientific papers. PageRank defines the ranking score of each node in a directed network, and the ranking score of node $i$, $s_i$, should satisfy:

\begin{equation}
\label{eq.pr}
s_i=c\sum_{j=1}^N\left(\frac{A_{ji}}{k_{j}^{out}}(1-\delta_{k_{j}^{out},0})+\frac{1}{N}\delta_{k_{j}^{out},0}\right)s_{j}+\frac{1-c}{N}.
\end{equation}
and $\|s\|=1$. In equation \ref{eq.pr}, $\delta_{a,b}=1$ when $a=b$ and $\delta_{a,b}=0$ otherwise, $N$ is the total number of nodes in the network, $k_{j}^{out}$ is the out degree of node $j$, and $A$ is the adjacency matrix of the network. The PageRank algorithm simulates the behaviour of ``random surfing" in the internet. While surfing randomly in the internet, people start their surfing by visiting a web page, and then they either follow one of the page's hyperlink to visit another web page or choose any other web pages in the internet randomly. The PageRank score captures the frequency of a particular web page to be visited by a random surfer, and the PageRank score of a node is divided among its forward links evenly to contribute to the PageRank scores of the nodes it point to. PageRank score can be computed by starting with any set of initial score and iterating the computation until it converges. $c$ is a damping factor controls the probability to follow a hyperlink while surfing. It can be set between 0 and 1. In case of citation networks, the value of $c$ is usually set to 0.5\cite{sciPR1,sciPR2}. Accordingly, we use $c=0.5$ in our study.

\subsubsection{CiteRank}

In PageRank, it assumes that a web surfer starts his/her surfing from a randomly selected web page. But it is unreasonable for the citation networks. There are aging effects in bibliometrics which can not be ignored. Researchers typically start ``surfing" scientiﬁc publications from a rather recent publication which caughts their attention. CiteRank\cite{citerank} (CR) is a more practical model for the diffusion along the citation networks. It is an adaptation of PageRank algorithm to citation networks. The CiteRank score represents the recent popularity of a scientific paper and can be used to predict the popularity of the paper. It assumes that researchers preferentially start from a rather recent paper and progressively follow the paper's references with every step while ``surfing" the citation network. The traffic of each paper that is calculated by CiteRank algorithm is ultimately used to predict the paper's popularity. CiteRank defines transfer matrix $W$ associated with the citation network as:
\begin{equation}
\label{eq.crw}
w_{ij}=\left\{
\begin{array}{lcl}
\frac{1}{k_{j}^{out}}  , &\mbox{if $j$ cites $i$}\\
0  , & otherwise
\end{array}
\right.
,
\end{equation}
where $k_{j}^{out}$ is the out-degree of the paper $j$. The probability of initially selecting the paper $i$ to start a random work is:
\begin{equation}
\label{eq.crp}
\rho_i=e^{-age_i/\tau},
\end{equation}
where $\tau$ is the decay parameter and $age_i$ is the age of paper $i$. Using $\vec{\rho}$ to express the vector of probability defined by Eq.\ref{eq.crp}, and the probability of encountering each paper after following one link can be calculated with $\alpha W\vec{\rho}$. The CiteRank score of each paper is then defined as the probability of encountering it via paths of any length:

\begin{equation}
\label{eq.cr}
\vec{S}=\vec{\rho}+\alpha W\vec{\rho}+\alpha^2W^2\vec{\rho}+\alpha^3W^3\vec{\rho}+\cdots.
\end{equation}
$\alpha$ is the probability that a researcher will follow a reference to an adjacent paper. CiteRank score is calculated by taking successive terms in above expansion to sufficient convergence.

\subsubsection{Rescaled PageRank}
PageRank has a strong bias towards the old papers which have accumulated many citations, so it is hard to predict the future influence of newly published papers which have not get many citations. The rescaled PageRank\cite{medo2016ji} (RS) uses z-score to factor out the bias caused by age in PageRank. The rescaled PageRank score of a paper $i$ is calculated by the z-score of PageRank score for a group of papers published in a similar time as $i$:

\begin{equation}
\label{eq.rs}
s_i=\frac{p_i-\mu_i}{\sigma_i},
\end{equation}
where $\mu_i$ and $\sigma_i$ are respectively the mean and standard deviation of the PageRank score of the group, and $p_i$ is the PageRank score of the paper $i$. Ordering papers by age, the group of papers used to calculate the rescaled PageRank score of paper $i$ are the papers $j\in[i-\Delta_p/2, i+\Delta_p/2]$. The parameter $\Delta_p$ represents the number of papers in the average window of each paper. $\Delta_p$ is set as 1000 in this paper. The rescaled PageRank score larger than zero indicates the paper out-performs other papers of similar age.

\subsection{Age-based diffusion model}

Both PageRank and CiteRank models are based on diffusive random walk. They do not exactly agree with the real process on citation networks.
First, they only select one node to follow at each diffusion step, while researchers may select several references of a paper to follow. Second, they do not take into account the correlation between steps, however the probability that a researcher will follow a paper's references is decay with the increase of steps. So, in this paper, we propose an age-based diffusion (AD) model which employs a process that analogous to random walk across the citation networks. The probability of initially selecting the paper $i$ to start a random work is exponentially decay with the paper's age, which is the same with CiteRank model defined in equation (\ref{eq.crp}). Considering the aging effects in bibliometrics, we introduce an exponential decay mechanism while distributing a paper's ranking score to its references. Instead of dividing among a paper's forward links evenly to contribute to the ranking scores of its references, a paper's ranking score is directly distributed to its references in the AD model. So the transfer matrix $W$ of the AD model is defined as:
\begin{equation}
\label{eq.adw}
w_{ij}=\left\{
\begin{array}{lcl}
 e^{-age_j/\tau} , &\mbox{if $j$ cites $i$}\\
0  , & \mbox{otherwise}
\end{array}
\right.
,
\end{equation}
where $\tau$ is the same decay parameter used in equation (\ref{eq.crp}) and $age_j$ is the age of paper $j$. The ranking score vector of the papers is then defined as:
\begin{equation}
\label{eq.ad}
\vec{S}=\vec{\rho}+\alpha_1 W\vec{\rho}+(\prod_{i=1}^{2}\alpha_i)W^2\vec{\rho}+(\prod_{i=1}^{3}\alpha_i)W^3\vec{\rho}+\cdots.
\end{equation}
where $\vec{\rho}$ is the vector of probability that defined by equation (\ref{eq.crp}) and $\alpha_i$ is the probability that a researcher will follow the references to adjacent papers at the $ith$ step. This probability that a researcher will follow a paper's references should decay with steps, because a researcher will be distracted rapidly while following paper's reference step by step. So $\alpha_i$ is set as a decreasing sequence:
\begin{equation}
\label{eq.ada}
\alpha_i = \alpha/{10^{i-1}}.
\end{equation}
$\alpha$ is the probability that a researcher will follow a reference to an adjacent paper at the first setp. The ranking score of each paper can be get by equation (\ref{eq.ad}) with enough terms to convergence.

In the AD model, the ratio of the ranking score that a paper will distribute to all its references exponentially decays with the paper's age, which is consistent with the real condition that the probability to follow an old paper's reference is small. And the probability to follow references decays with the increase of steps, which is consistent with the real condition that a researcher will be distracted rapidly while following paper's references step by step. By introducing this two mechanisms, the AD model we propose in this paper can better simulate the attention diffusion process in citation networks.

\section{Results}

The popularity prediction problem we studied in this paper is to identify which paper can receive more citations in the near future and will be popular. We use the APS data set introduced in section 2.1 to test the performance of different prediction methods. While evaluating the prediction performance of different prediction methods, the citation data before the testing time $t$ is used as training set to calculate the prediction score, and the citation date in a future time window $T_f$ just after the time $t$ is used as test set to evaluate the performance. For a given time $t$, the citation increase of paper $i$ in future $T_f$ time window is
\begin{equation}
\label{eq.tf}
f_i(t)=\Delta c_i(t,T_f)=c_i(t+T_f)-c_i(t),
\end{equation}
where $c_i(t)$ is the number of citations paper $i$ receives until time $t$. The quantity $f_i(t)$ can measure the temporal interest in paper $i$, and can be considered as the real future popularity of paper $i$ after time $t$. The age of paper $i$ is calculated by the difference between the publication date of the paper and the testing time $t$.

As all the prediction methods we considered in this paper are based on papers' citation, we removed the papers that have not get any citations before the testing time $t$ from the training set and test set.

\subsection{Evaluation metrics}

The prediction score calculated based on the training set could be compared with the real future popularity calculated based the test set to evaluate the performance of a prediction method. For a given paper $i$, we use $s_i$ to denote its prediction score and $f_i$ to denote its real future popularity. We introduced three metrics to evaluate the prediction performance of different methods in this paper: Pearson correlation coefficient, Spearman rank correlation coefficient and precision.

Pearson correlation coefficient is a good way to quantify the similarity between two lists. The Pearson correlation coefficient between the prediction score and the real future popularity of papers is given by
\begin{equation}
\label{eq.pc}
r=\frac{1}{N}\sum\left(\frac{s_i-\bar{s}}{\sigma_s}\right)\left(\frac{f_i-\bar{f}}{\sigma_f}\right),
\end{equation}
where $\sigma_s$ and $\sigma_f$ are, respectively, the standard deviations of the prediction score and the real future popularity of papers, and $\bar{s}$ and $\bar{f}$ are their expected values. The values of Pearson correlation coefficient are in the range $[-1,1]$. A higher value indicates a higher correlation between two lists.

While evaluating the performance of a prediction method, the prediction score can be mapped into a predicted ranking, and then we could compare the predicted ranking with the real ranking. Spearman rank correlation coefficient is simply a special case of the Pearson correlation coefficient in which two sets of data are converted to rankings before calculating the coefficient. We use $r_{s_i}$ to denote the ranking of the paper $i$ ranked by prediction score and $r_{f_i}$ to denote the ranking of the paper $i$ ranked by real future popularity. If there are equal values in the ranked list, they will be assigned the same ranking using the average of their positions. The Spearman rank correlation coefficient is calculated by
\begin{equation}
\label{eq.sc}
\rho=\frac{1}{N}\sum\left(\frac{r_{s_i}-\bar{r_s}}{\sigma_{r_s}}\right)\left(\frac{r_{f_i}-\bar{r_f}}{\sigma_{r_f}}\right),
\end{equation}
where $\sigma_{r_s}$ and $\sigma_{r_f}$ are, respectively, the standard deviations of the rankings by prediction score and the real future popularity of papers, and $\bar{r_s}$ and $\bar{r_f}$ are their expected values.

Instead of predicting the popularity of all papers, it is more important to predict which paper will be popular in the future. So when evaluating the performance of a prediction method, we can only consider the top part of the ranking list. Precision is a metric based on this. Precision calculates the proportion of papers that are ranked in the top $n$ positions by prediction score are really top $n$ papers ranked by real future popularity. The precision of a prediction method is defined as $P_n = D_n/n$ , where $D_n$ indicates the number of common papers in the top $n$ positions of the predicted ranking and the real ranking. It ranges from 0 to 1, the higher the better. In this paper, we restricted our analysis to top $1\%$ papers in the test set.

\subsection{The effectiveness of the AD model}

\begin{figure}[hbp]
\centering
\scalebox{1}[1]{\includegraphics{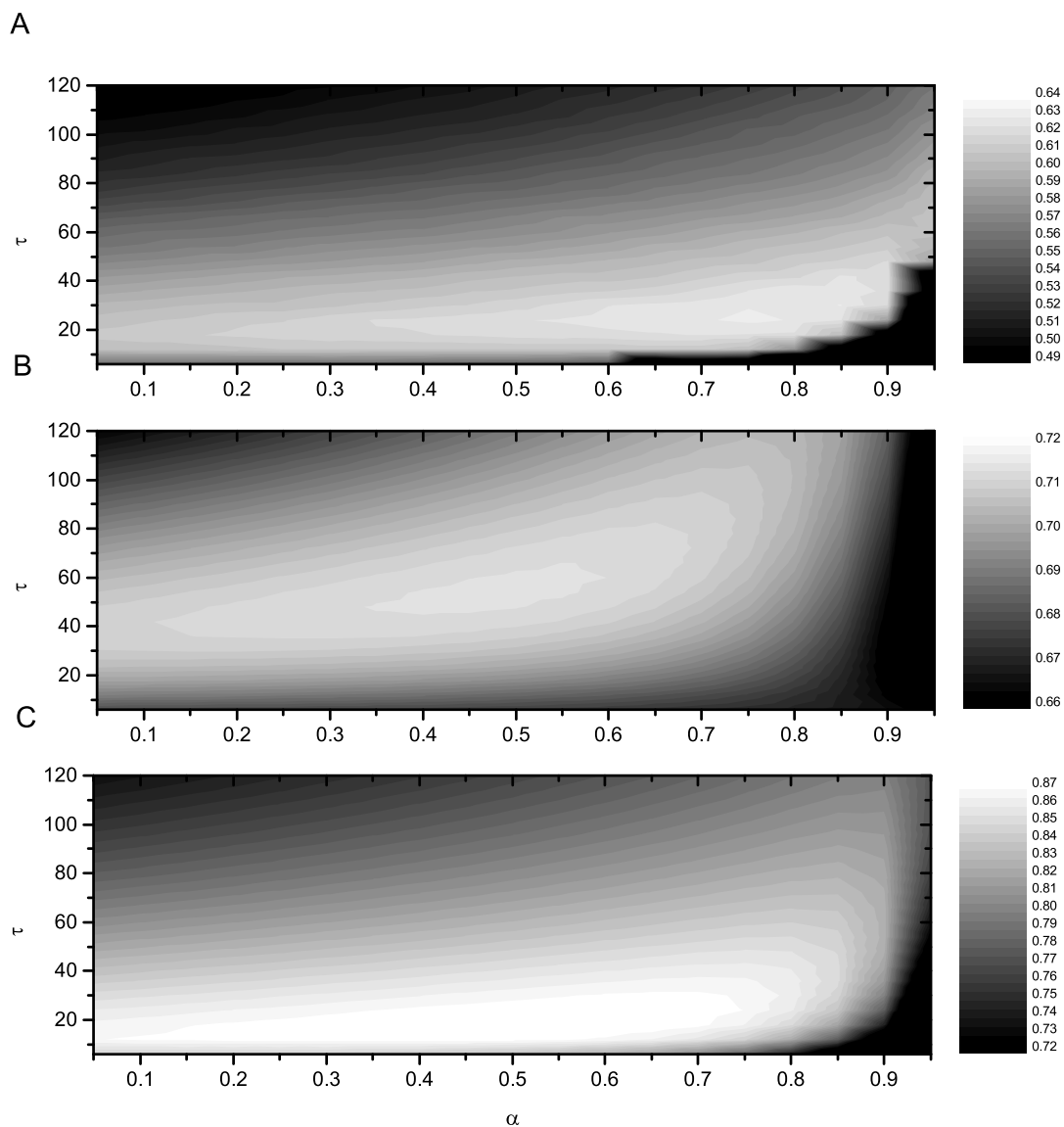} }\caption{
The performance of the AD model with different parameters $\tau$ and $\alpha$. (A) Precision. (B) The Spearman rank correlation coefficient. (C) The Pearson correlation coefficient. The testing time is set as the beginning of the year 2010, and the future time window $T_f$ is $5$ years.}
\label{fig1}
\end{figure}

To evaluate the performance of the AD model with different parameters $\tau$ and $\alpha$, we selected the testing time as the beginning of the year 2010 and the future time window $T_f$ as $5$ years. In this case, We used the data before 2010 to predict the popularity of papers, and used the data in a future time window of 5 years to test the performance. Figure \ref{fig1} shows the result. From figure \ref{fig1}, we could find that, for a given $\alpha$, a suitable decay parameter $\tau$ can significantly increase the prediction performance. This indicates that papers' temporal information plays an important role in predicting papers' popularity. The best performance of AD model can be achieved by a suitable value of $\tau$ and $\alpha$. The highest precision is achieved when $\alpha=0.74$ and $t=24$, the highest Spearman rank correlation coefficient is achieved when $\alpha=0.49$ and $t=54$, and the highest Pearson correlation coefficient is achieved when $\alpha=0.54$ and $t=18$. As the three evaluation metrics focus on different aspects, there is an appreciable discrepancy in the best parameters of the three evaluation metrics.

\begin{figure}[htbp]
\centering
\scalebox{0.7}[0.7]{\includegraphics{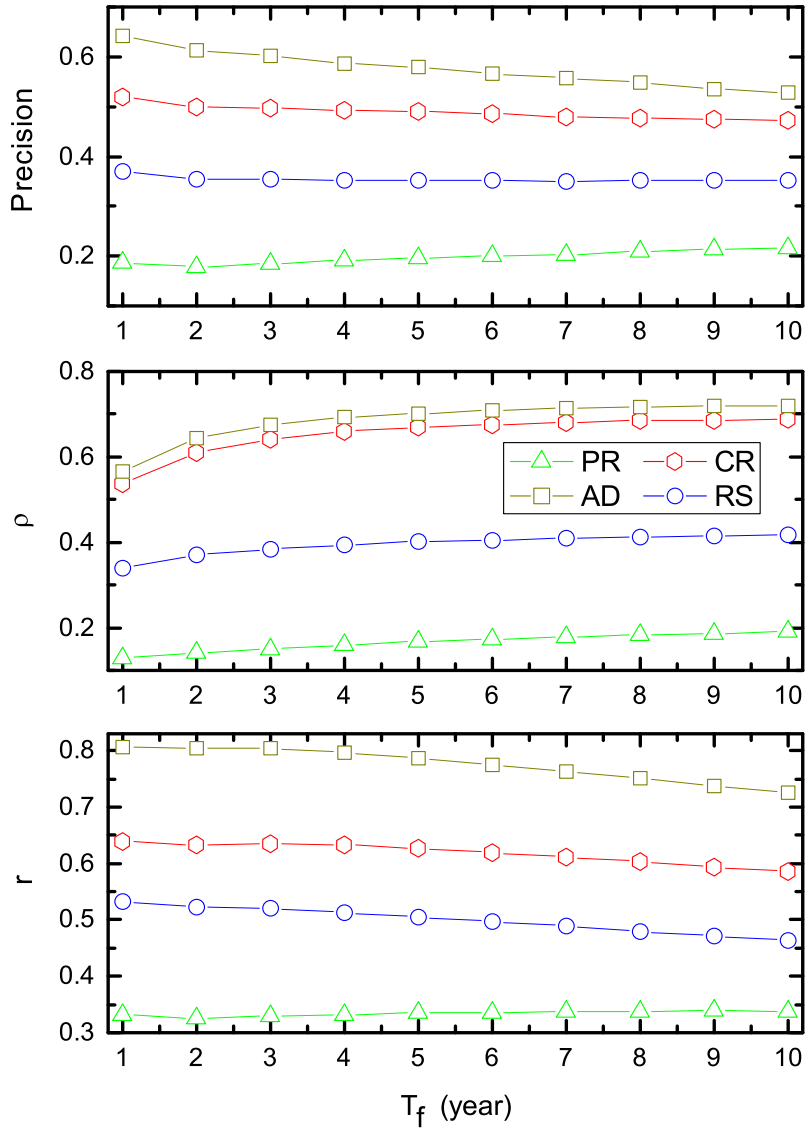} }\caption{
The prediction performance PageRank (PR), CiteRank (CR), rescaled PageRank(RS) and age-based diffusion (AD) model under different $T_f$. The results are obtained by averaging the results of $5$ randomly select testing time $t$.}
\label{fig2}
\end{figure}

While evaluating the performance of a prediction method, a small future time window $T_f$ represents a short term popularity, and a large $T_f$ represents a long term popularity. Therefore, we tested the performance of different prediction methods under different $T_f$ to compare their performance. To eliminate the influence that may be caused by different testing time, we obtained the final results by averaging the results of five randomly selected testing time $t$. To make sure that there is enough data in both the training set and the test set, the five randomly selected testing time $t$ is select in the range from January 1990 to January 2006. The parameters for CiteRank and the AD model are selected as the one corresponding to  the highest value of the target metric. Figure \ref{fig2} shows the prediction performance of different prediction method as a function of the future time window $T_f$.

From figure \ref{fig2}, it is clear that the AD model outperforms other methods on all the three evaluation metrics for all $T_f$ value. The performance of PageRank is significantly worse than other methods. As PageRank model dose not take any temporal information while calculating the result, the prediction performance of PageRank may be limited. For CiteRank, rescaled PageRank, and the AD models, the precision and the Pearson correlation coefficient have a small decrease with the increase of $T_f$. This indicates that it becomes more difficult to predict a longer term popularity which may be influenced by more unexpected factors. But the Spearman rank correlation coefficient increases with $T_f$. The reason may be that too many papers have the same real popularity value but different prediction score when $T_f$ is small, which can decrease the Spearman rank correlation coefficient while comparing the difference between the two ranking list.

\begin{figure}[htbp]
\centering
\scalebox{1}[1]{\includegraphics{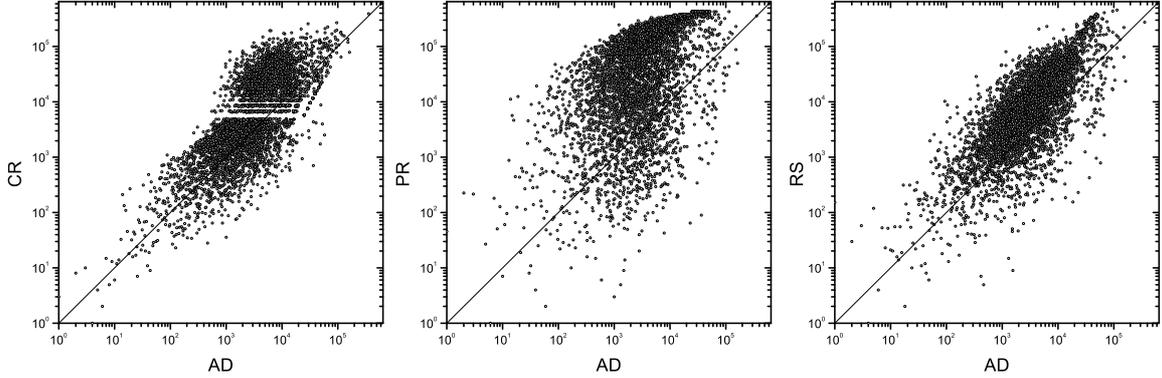} }\caption{
The comparison of ranking results of the real top $1\%$ papers. The x-axis is the ranking obtained by age-based diffusion (AD) model, and the y-axis is the ranking obtained by CiteRank (CR), PageRank (PR) and rescaled PageRank (RS) respectively. Black solid lines in all figures are $y=x$, indicating that the dot on it has the same rankings in the two methods.}
\label{fig3}
\end{figure}

\subsection{Top Papers Analysis}

In reality, people is more conscious of the papers that will be popular in the future rather than the real future popularity of all papers. So, we analyzed the ranking results of different methods for the papers that are ranked as top $1\%$ papers. We selected the testing time as the beginning of the year 2010 and the future time window $T_f$ as $5$ years in this subsection to analysis the results. The parameters for CiteRank and the AD model are selected as the one corresponding to  the highest value of precision.

Figure \ref{fig3} shows the ranking results of CiteRank, PageRank and rescaled PageRank methods compared with the ranking result of the AD model for the papers that are ranked as top $1\%$ papers by real future popularity. Black solid lines are $y=x$. If a dot is positioned up the solid lines, it means that the paper represented by the dot is ranked lower by the corresponding method than the AD model. It is clear that major parts of the dots are located up the $y=x$ lines in all the three graphs. These indicate that, the AD model can rank the real top $1\%$ papers higher than other three methods.

\begin{figure}[htbp]
\centering
\scalebox{0.8}[0.8]{\includegraphics{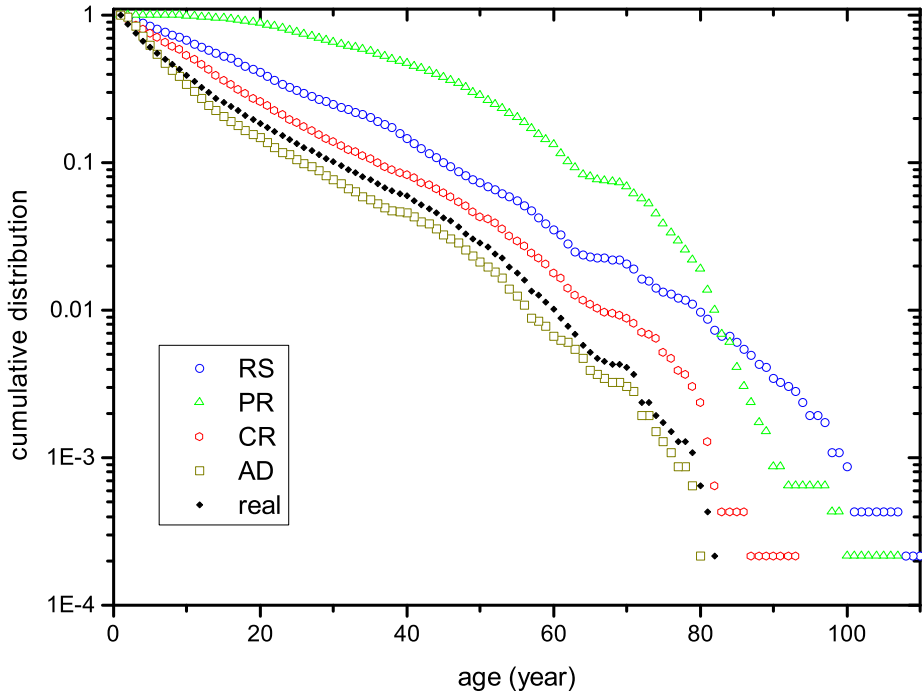} }\caption{
The cumulative distribution of the age of the top $1\%$ papers ranked by different methods. The results of CiteRank (CR), PageRank (PR), rescaled PageRank (RS), and the age-based diffusion (AD) model are shown respectively. The real top $1\%$ papers that ranked by the data in the future time window are also shown for comparison.}
\label{fig5}
\end{figure}

Figure \ref{fig5} shows the cumulative distribution of the age of the top $1\%$ papers ranked by different methods. The cumulative distribution of the real top $1\%$ papers ranked by date in the future time window exponentially decays with age. This indicates that the number of papers ranked in the real top $1\%$ decays rapidly with age, and only a small number of those papers are old papers. The cumulative distribution of the AD model has a similar exponential distribution as the real result. The distribution of CiteRank, PageRank, rescaled PageRank methods have an obvious deviation from the real result, and many old papers are ranked as top $1\%$ papers.

\begin{figure}[htbp]
\centering
\scalebox{0.8}[0.8]{\includegraphics{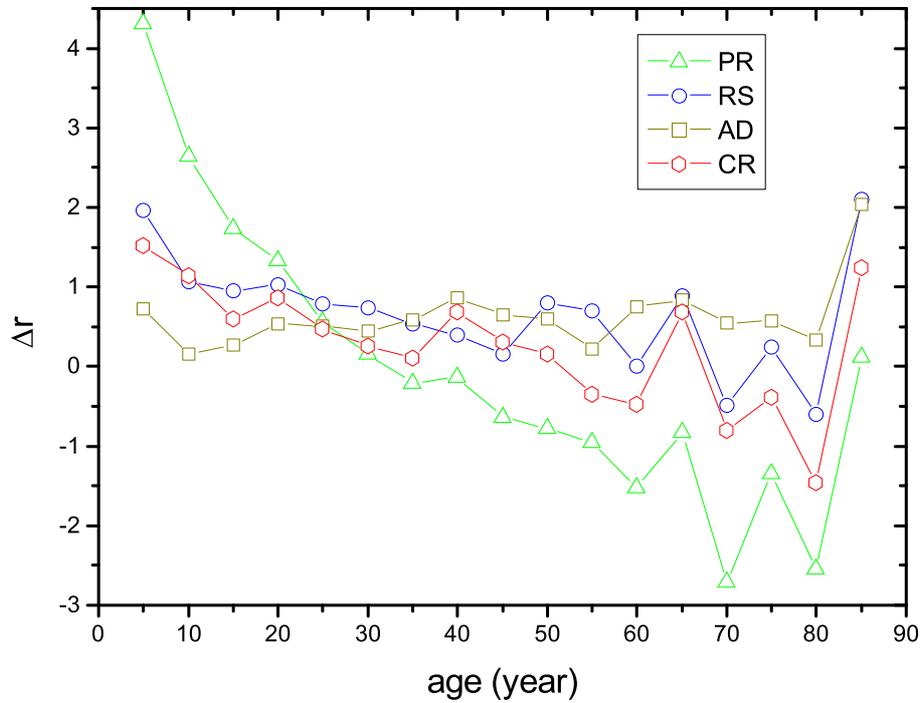} }\caption{
The average value of $\Delta r$ for the papers under different age. The results of CiteRank (CR), PageRank (PR), rescaled PageRank (RS), and the age-based diffusion (AD) model are shown respectively. Only the top $1\%$ papers in the future time window are considered to show the ranking difference of the top papers with different age.}
\label{fig4}
\end{figure}

To better understand the age bias of different methods, we analyzed the ranking difference of different methods for the real top $1\%$ papers with different age. We set the ranking difference of a paper as $\Delta r=log(r_p)-log(r_f)$, where $r_f$ is the paper's real ranking in the future and $r_p$ is the paper's ranking by a prediction method. We take the logarithm of the rankings to avoid the influence of the lowly ranked papers while calculating the average result. $\Delta r=0$ indicates the predicted ranking is echo to the real ranking, and the prediction method has a perfect performance; $\Delta r >0$ indicates the prediction method underestimate the paper's popularity. $\Delta r < 0$ indicates the prediction method overrate the paper's popularity. Figure \ref{fig4} shows the average value of $\Delta r$ for the papers under different age. This figure only considers the real top $1\%$ papers in the future time window. Figure \ref{fig4} shows that, PageRank, CiteRank, and rescaled PageRank methods have an obvious bias for papers with different age. Newly published papers are underestimated and very old papers are overrated. While the AD model has no obvious bias for papers with different age, and the average $\Delta r$ value is close to zero for all ages. This indicates that the AD model has a good prediction performance for both old and new papers.

\begin{figure}[htbp]
\centering
\scalebox{1}[1]{\includegraphics{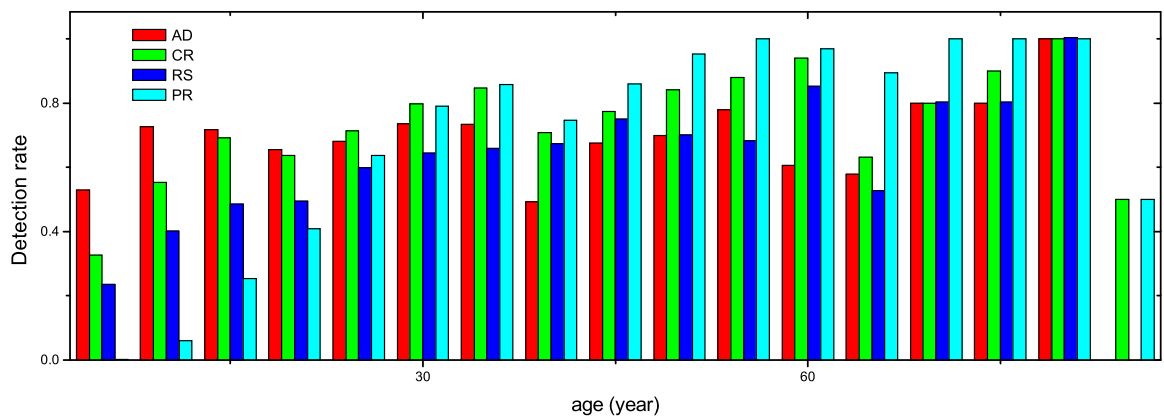} }\caption{
The correlation of detection rate with age under different prediction methods. The results of CiteRank (CR), PageRank (PR), rescaled PageRank (RS), and the age-based diffusion (AD) model are shown respectively.}
\label{fig6}
\end{figure}

A prediction method may have different precisions for papers with different ages. We separated the real top $1\%$ papers in bins by age of every 5 years. The detection rate of each bin is defined as the rate of papers that can be detected as top $1\%$ papers by a prediction method. We plotted figure \ref{fig6} to show the correlation of detection rate with age under different prediction methods. It is obviously that, PageRank can only detect very old papers as top papers, and the detection rate of newly published papers is nearly zero. The rescaled PageRank can factor out the bias that caused by age in PageRank, and the detection rate of newly published papers is obviously higher than the result of PageRank. CiteRank can further improve the detection rate of newly published papers, while keeping a relatively high detection rate of old papers. Compared with CiteRank, the AD model highly increases the detection rate of the latest papers that less than 10 years old but decreases the detection rate of papers that older than 30 years. We have found in figure \ref{fig5} that the number of papers ranked in the real top $1\%$ is decay rapidly with age, and most of the top papers are newly published papers. So the detection rate of newly published papers is more important than the detection rate of old  papers, and a small increase in the detection rate of newly published papers may result in a big improvement of the total prediction precision. The result is that the AD model has the highest overall precision.

From the experimental result, we could find that, AD model has a good ability to predict which newly published papers will be popular in the near future while keeping a relatively high overall prediction precision.

\section{Conclusion and Discussion}

The age-based diffusion (AD) model we propose in this paper is a model simulating the attention diffusion process along the citation networks. Researchers typically start ``surfing" scientific publications from a rather recent publication that catch their attention, and then they will follow the paper's references with the probability that is exponentially decaying with the paper's age and also decays with the iteration of steps. The AD model is a time-aware method, and we use it to predict the future popularity of scientific papers. We introduced three metrics to evaluate the performance in citation prediction: Pearson correlation coefficient, Spearman rank correlation coefficient and precision. The experimental results show that the AD model can improve the prediction accuracy compared with other benchmark methods. For some newly published papers that have not accumulated many citations but will be popular in the near future, the AD model can substantially improve their rankings. This is critical, because identifying the future highly cited paper from thousands of new papers published each month would provide very valuable references for researchers.

While predicting the future popularity of scientific papers using citation networks, we have to face the "cold start" problem. This problem refers to the case in which newly published papers have zero or few citations. Time-aware methods may increase the possibility to detect the potential popular papers that are newly published. But essentially, implementing time-aware methods still need citation information, and such kind of methods have inherent limitations to predict the popularity of newly published papers that have zero or few citations. Exploiting paper metadata is one way to solve the ``cold start" problem. Although metadata based prediction methods do not seem to have significant advantage in popularity prediction of papers\cite{TKDE2019}, combining network based prediction methods with metadata based methods may eventually solve the ``cold start" problem and have a much higher overall predict precision. This will be the direction of our future works on popularity prediction of scientific papers.

\section*{Acknowledgements}

We are grateful to An Zeng for useful comments and suggestions. This work was supported in part by the National Nature Science Foundation of China under Grant 61603340 and 61773348, and in part by the National Nature Science Foundation of Zhejiang Province under Grant LY19F020023.

\section*{References}




\begin{thebibliography}{00}

\bibitem{plos2010}
    Fanelli D. Do Pressures to Publish Increase Scientists' Bias? An Empirical Support from US States Data. PLOS ONE, 2010, 5(4): e10271.

\bibitem{nature2016}
    Sarewitz D. The pressure to publish pushes down quality. Nature, 2016, 533(7602):147-147.

\bibitem{review2017pr}
    Zeng A, Shen Z, Zhou J, et al. The science of science: From the perspective of complex systems. Physics Reports, 2017, 714: 1-73.

\bibitem{review2015}
    Mingers J, Leydesdorff L. A review of theory and practice in scientometrics. European journal of operational research, 2015, 246(1): 1-19.

\bibitem{review2016ji}
    Waltman L. A review of the literature on citation impact indicators. Journal of informetrics, 2016, 10(2): 365-391.

\bibitem{hindex}
    Hirsch J E. An index to quantify an individual's scientific research output. Proceedings of the National academy of Sciences, 2005, 102(46): 16569-16572.

\bibitem{if}
   Garfield E. The history and meaning of the journal impact factor. Jama, 2006, 295(1): 90-93.

\bibitem{paperpa}
  Price D S. A general theory of bibliometric and other cumulative advantage processes. Journal of the American society for Information science, 1976, 27(5): 292-306.

\bibitem{newman2009}
    Newman M E J. The first-mover advantage in scientific publication. Europhysics Letters, 2009, 86(6): 68001.

\bibitem{newman2014}
    Newman M E J. Prediction of highly cited papers. Europhysics Letters, 2014, 105(2): 28002.

\bibitem{medo2016ji}
    Mariani M S, Medo M, Zhang Y C. Identification of milestone papers through time-balanced network centrality. Journal of Informetrics, 2016, 10( 4):1207-1223.

\bibitem{medo2017ji}
    Vaccario G, Medo M, Wider N, et al. Quantifying and suppressing ranking bias in a large citation network. Journal of Informetrics, 2017, 11( 3):766-782.

\bibitem{fitness1}
    Caldarelli G, Capocci A, De Los Rios P, et al. Scale-free networks from varying vertex intrinsic fitness. Physical review letters, 2002, 89(25): 258702.

\bibitem{fitness2}
  Bianconi G, Barab{\'a}si A L. Competition and multiscaling in evolving networks. Europhysics Letters, 2001, 54(4): 436.

\bibitem{medoprl}
   Medo M, Cimini G, Gualdi S. Temporal effects in the growth of networks. Physical review letters, 2011, 107(23): 238701.

\bibitem{zhou2020}
    Zhou Y, Cheng H, Li Q, Wang W. Diversity of temporal influence in popularity prediction of scientific publications, Scientometrics, 2020, 123(1): 383-392.

\bibitem{adecay2015}
      Parolo P D B, Pan R K, Ghosh R, et al. Attention decay in science. Journal of Informetrics, 2015, 9(4): 734-745.

\bibitem{science2013}
  Wang D, Song C, Barab{\'a}si A L. Quantifying long-term scientific impact. Science, 2013, 342(6154): 127-132.

\bibitem{zhou2011}
    Zhou Y B , L{\"u} L, Li M. Quantifying the influence of scientists and their publications: distinguishing between prestige and popularity[J]. New Journal of Physics, 2011, 14(3):33033-33049(17).

\bibitem{pagerank}
    Page L. The pagerank citation ranking: Bringing order to the web. Stanford InfoLab, 1999, 1-14.

\bibitem{anatomy}
    Brin S, Page L. The anatomy of a large-scale hypertextual Web search engine. Computer Networks \& Isdn Systems, 1998, 30.

\bibitem{sciPR1}
    Chen P, Xie H, Maslov S, Redner S. Finding scientific gems with Google’s PageRank algorithm. Journal of Informetrics, 2007, 1(1): 8-15.

\bibitem{sciPR2}
    Ma N, Guan J, Zhao Y. Bringing PageRank to the citation analysis. Information Processing \& Management, 2008, 44(2): 800-810.

\bibitem{citerank}
    Walker D, Xie H, Yan K K, Maslov S. Ranking scientific publications using a model of network traffic. Journal of Statistical Mechanics: Theory and Experiment, 2007, 2007(06):P06010-P06010.

\bibitem{rpagerank}
    Mariani M S, Medo M, Zhang Y C, Identification of milestone papers through time-balanced network centrality, Journal of Informetrics, 2016,10(4), 1207-1223,

\bibitem{TKDE2019}
    Kanellos I, Vergoulis T, Sacharidis D, et al. Impact-Based Ranking of Scientific Publications: A Survey and Experimental Evaluation. IEEE Transactions on Knowledge and Data Engineering, 2019: 1-1.

\end{thebibliography}


\end{document}